\documentstyle[aps,prl,epsfig,rotating,twocolumn]{revtex}

\begin{document}

\twocolumn[\hsize\textwidth\columnwidth\hsize\csname
@twocolumnfalse\endcsname
\title{Mirages, anti-mirages, and further surprises in quantum corrals with 
 non-magnetic impurities}
\author{Markus~Schmid and Arno~P.~Kampf }
\address{Theoretical Physics III, Center for Electronic Correlations
and Magnetism,  Institute of Physics, University of Augsburg,
86135 Augsburg, Germany}
\date{July 22, 2003}
\maketitle
\begin{abstract}
We investigate the local density of states (LDOS) for non-interacting 
electrons in a hard-wall ellipse in the presence of a single non-magnetic
scattering center. Using a T-matrix analysis we calculate the local Green's 
function and observe a variety of quantum mirage effects for different 
impurity positions. Locating the impurity near positions with LDOS maxima 
for the impurity free corral can either lead to a reduction or an ehancement 
of the LDOS at the mirror image point, i.e. a mirage or anti-mirage effect, or
even suppress LDOS maxima in the entire area of the corral. 

\end{abstract}

 ]
\section{Introduction}
The technological advances in scanning tunneling microscopy (STM) have made 
it possible to manipulate individual atoms on metallic surfaces \cite{Eigler}. 
These remarkable achievements allow local measurements of electronic 
properties with a spatial resolution of atomic length scales even on 
artificially designed geometries for surface adatoms \cite{Fiete}. The 
experimental advances have led e.g. to the spectacular observation of 
mirage effects in elliptic quantum corrals of magnetic Co atoms on Cu (111) 
surfaces \cite{Manoharan}. If an additional Co atom is placed at one of the 
foci of the ellipse, a Kondo resonance in the local density of states (LDOS) 
is not only observed at the magnetic ion itself but at the other, impurity 
free focus as well. The mirage effect can be viewed as a beautiful 
manifestation of quantum mechanical interference phenomena as a result of the 
multiple scattering events of the electrons from the impurity and the atoms 
forming the boundary of the corral and from the impurity at one focus.

The theoretical work on this problem emphasised the many-body Kondo physics 
of itinerant electrons in the corral interacting with the localized magnetic 
moments of the impurities \cite{Schiller,Lobos,Aligia,Chiappe}. But it was 
also found that for the specific elliptic geometry even stronger mirage 
effects can be observed, if the additional impurity is moved slightly away 
from the focal point \cite{Aligia}. Furthermore, mirage effects may even be 
absent depending on the intrinsic level width \cite{Lobos} or for specific 
electron densities inside the elliptic corrals; one of the crucial quantities 
is the LDOS at the position where the additional adatom is placed. Therefore, 
already without invoking the Kondo physics the unique geometry of an elliptic 
corral appears to give rise to intriguing quantum mechanical effects which 
deserve a detailed analysis also for the significantly 
simpler problem of a non-magnetic impurity in a hard-wall corral.

Indeed we will show in this paper that a surprising variety of structures in 
the LDOS can be generated depending on the precise location of a single 
additional non-magnetic scattering center inside a hard-wall elliptic corral. 
At the mirror position of the local impurity potential the LDOS may be 
suppressed or even enhanced, and also the almost complete suppression of LDOS 
maxima can be achieved. These observations underline the richness of quantum 
mechanical interference phenomena in an elliptical geometry which may be 
verified experimentally.

\section {Solution of the 2d Schr\"odinger Equation in a Hard Wall Ellipse}
\noindent In a first step we review the solution of the two-dimensional (2D) 
Schr\"odinger equation for non-interacting electrons in a hard wall ellipse. 
Following Ref. \cite{Porras} it is convenient to introduce elliptical 
coordinates through the transformation \cite{McLachlan}
\begin{eqnarray}
&& x = a e \cos(\theta) \cosh(\eta)\, , \nonumber \\ 
&& y = a e \sin(\theta) \sinh(\eta)\, ,
\label{Hamstart}
\end{eqnarray}
where $a$ denotes the semimajor axis and $e$ is the eccentricity of the 
ellipse. In
the new coordinates the Schr\"odinger equation takes the following form:
\begin{eqnarray}
\Big[&-&\frac{\hbar^2}{2m(ae)^2}\frac{2}{\cosh(2\eta)-\cos(2\theta)} 
\left(\frac{\partial^2}{\partial\theta^2}+\frac{\partial^2}{\partial\eta^2}
\right)\nonumber\\
&+& V(\eta)\Big]\psi(\theta,\eta)=\epsilon\psi(\theta,\eta)
\label{SchEq}
\end{eqnarray}
where $m$ is the electron mass and $\epsilon$ is the energy eigenvalue. Due to 
the hard-wall condition the potential $V$ vanishes inside the elliptic corral 
and is infinite otherwise. Using the factorized ansatz for the eigenfunctions
\begin{eqnarray}
&& \psi(\theta,\eta)=\Theta(\theta) \Phi(\eta)
\end{eqnarray}
the Schr\"odinger equation inside the ellipse ($V(\eta)=0$) is rearranged as 
\begin{equation}
-\left[\frac{\partial^2}{\partial\theta^2} - 2 k \cos(2\theta)\right]
\Theta(\theta) =  \left[\frac{\partial^2}{\partial\eta^2}
+2 k \cosh(2\eta) \right]\Phi(\eta)
\label{facSE}
\end{equation}
where $k=(ae)^2 m\epsilon/2\hbar^2$. The Schr\"odinger equation thus 
separates with respect to the elliptic coordinates $\theta$ and $\eta$ and 
reduces to the two differential equations 
\begin{eqnarray}
&&\frac{\partial^2}{\partial\theta^2}\Theta(\theta) 
+\left(\alpha - 2 k \cos(2\theta)\right)\Theta(\theta) = 0\, ,  
\label{sepSE1} \\
&& \frac{\partial^2}{\partial\eta^2}\Phi(\eta) -\left(\alpha - 2 k 
\cosh(2\eta) \right)\Phi(\eta)= 0
\label{sepSE2}
\end{eqnarray}
where $\alpha$ is a separation constant. Equations (\ref{sepSE1}) and 
(\ref{sepSE2}) are the Mathieu equation and the modified Mathieu equation,
 respectively \cite{McLachlan,Stegun}.

Since $\theta$ is the polar angle and thus $\Theta(\theta+2\pi)=\Theta
(\theta)$ we select the periodic solutions of Eq. (\ref{sepSE1}), which are
 the Mathieu functions of the first kind of integral order \cite{McLachlan}.
 We obtain two types of solutions:
\begin{eqnarray}
&& \Theta_{r}(\theta)=ce_r(\theta,k^c)
\label{Mathieu1a} \\
{\rm or}\hskip1.5cm && \Theta_{r}(\theta)=se_r(\theta,k^s)
\label{Mathieu1b}
\end{eqnarray}
where ``$ce$'' and ``$se$'' are the abbreviations for ``cosine-elliptic'' and 
``sine-elliptic'' introduced by Whittaker\cite{McLachlan}. Due to the 
periodicity condition there exists only a discrete set of numbers 
$\alpha=\alpha_r(k)$ for fixed $k$ \cite{Stegun}. The index $r$ in Eqs. 
(\ref{Mathieu1a}) and (\ref{Mathieu1b}) denotes the order of the Mathieu 
functions of the first kind. 

The solutions of Eq. (\ref{sepSE2}), which are necessarily restricted to the
 same set of numbers $\alpha_r(k)$ as the corresponding solution of 
(\ref{sepSE1}), are
\begin{eqnarray}
&& \Phi_r(\eta)= Ce_r(\eta,k^c)\, ,
\label{Mathieu2a}\\
{\rm or}\hskip1.5cm && \Phi_r(\eta)= Se_r(\eta,k^s)\, .
\label{Mathieu2b}
\end{eqnarray}
$Ce$ and $Se$ in Eqs. (\ref{Mathieu2a}) and (\ref{Mathieu2b}) denote the 
modified Mathieu functions of the first kind of integral order 
\cite{McLachlan}. They must meet the hard-wall condition $\Phi_r(\eta_0)=0$ at
 the boundary line of the ellipse, which leads to a discrete set of values
 $k_n$. This implies that the discrete sets for $k^c_n$ and $k^s_n$ must meet
 the conditions
\begin{eqnarray}
&& \Phi_r(\eta_0)= Ce_r(\eta_0,k^c_n)\, ,\\
{\rm or}\hskip1.5cm && \Phi_r(\eta_0)= Se_r(\eta_0,k^s_n)\, .
\end{eqnarray}
Therefore, $k^c_n$ and $k_n^s$ are the $n^{th}$ zeroes of $Ce_r(\eta_0,k^c)$ 
and $Se_r(\eta_0,k^s)$, respectively, for fixed $r$. $\eta_0$ is directly 
related to the eccentricity $e$ of the ellipse through
\begin{eqnarray}
&& e = \frac{1}{\cosh(\eta_0)}\, .
\label{ecc}
\end{eqnarray} 
Altogether we find the following form of the exact eigenstates inside the 
hard-wall ellipse \cite{Porras}:
\begin{eqnarray}
&& \psi_{r,n_c}^c(\theta,\eta) = ce_r(\theta,k_n^c) Ce_r(\eta,k_n^c)\, , 
\label{eigenstates1} \\
&& \psi_{r,n_s}^s(\theta,\eta) = se_r(\theta,k_n^s) Se_r(\eta,k_n^s) \, .
\label{eigenstates2}
\end{eqnarray} 
$r$, $n$ and $c(s)$ enumerate the quantum numbers for the eigenstates; the 
eigenenergies are determined by the hard-wall condition and the periodicity 
requirement for the Mathieu functions. Note that $ce$, $se$ and $Ce$, $Se$ are 
all real. The symmetry under reflection at the semimajor axis is even for 
the $ce_r$ function (cosine-elliptic) and odd for the $se_r$ function 
(sine-elliptic) for each $r$. Moreover the parity of $ce_r$ and $se_r$
 is even/odd, if $r$ is even/odd. Specifically,
\begin{eqnarray}
\psi_r^c(-\theta,\eta) & = & \psi_r^c(\theta,\eta)\ , \\
\psi_r^s(-\theta,\eta) & = & -\psi_r^s(\theta,\eta)\ , \\
\psi_{2r}^{c/s}(\theta+\pi,\eta) & = & \psi_{2r}^{c/s}(\theta,\eta)\, ,\\
\psi_{2r+1}^{c/s}(\theta+\pi,\eta) & = & -\psi_{2r+1}^{c/s}(\theta,\eta)\, .
\end{eqnarray}

In the following we will specifically investigate an ellipse with eccentricity 
$e=0.5$. The energy scale for the eigenenergies is furthermore determined by 
the size of the ellipse, i.e. the length of the semimajor axis $a$. For a 
specific choice for $a$ we orient ourselves at the typical size of the 
elliptic corral of Co atoms on a Cu (111) surface used in the experiments of 
Manoharan et al. in Ref. \cite{Manoharan} for which $a\sim 71$\AA. In this 
setup the Fermi energy $\epsilon_F$ is 450 meV which corresponds to the 
energy of the 42$^{nd}$ eigenstate, and due to the spin degeneracy amounts 
to a particle number of 84 electrons inside the ellipse. For the specific 
choice for the size of the ellipse the ground state energy 
\begin{eqnarray}
E_0 = k_0 \frac{2 \hbar^2}{(ae)^2 m}
\end{eqnarray}
is $E_0 = 13$ meV.
For fixed eccentricity $e=0.5$ and $k_n=k_{42}$ the wavefunction of the 
eigenstate at the Fermi surface has a high probability density at the two foci 
of the ellipse -- a precondiction to observe strong mirage effects when 
an additional impurity atom is placed at one of the focal points.
 
\section{The scattering problem} 
\noindent The LDOS of the non-interacting electron system inside the corral 
is easily obtained from the retarded one-particle Green's function 
\begin{eqnarray}
&& G_0^{ret}({\bf r,r'},\epsilon) = \sum_j \frac{\psi_j({\bf r}) 
   \psi_j^{\ast}({\bf r'})}{\epsilon - \epsilon_j + i \delta}
\end{eqnarray}
where the eigenfunctions $\psi_j$ are given by Eqs. (\ref{eigenstates1}) and
 (\ref{eigenstates2}) and $\epsilon_j$ is the corresponding eigenenergy of the
 j$^{th}$ eigenstate. The free LDOS $N_0({\bf r},\epsilon)$ then follows from 
\onecolumn
\begin{figure}[t]
\noindent
\begin{minipage}[t]{.46\linewidth}
\epsfxsize=13cm \centerline{\epsfbox{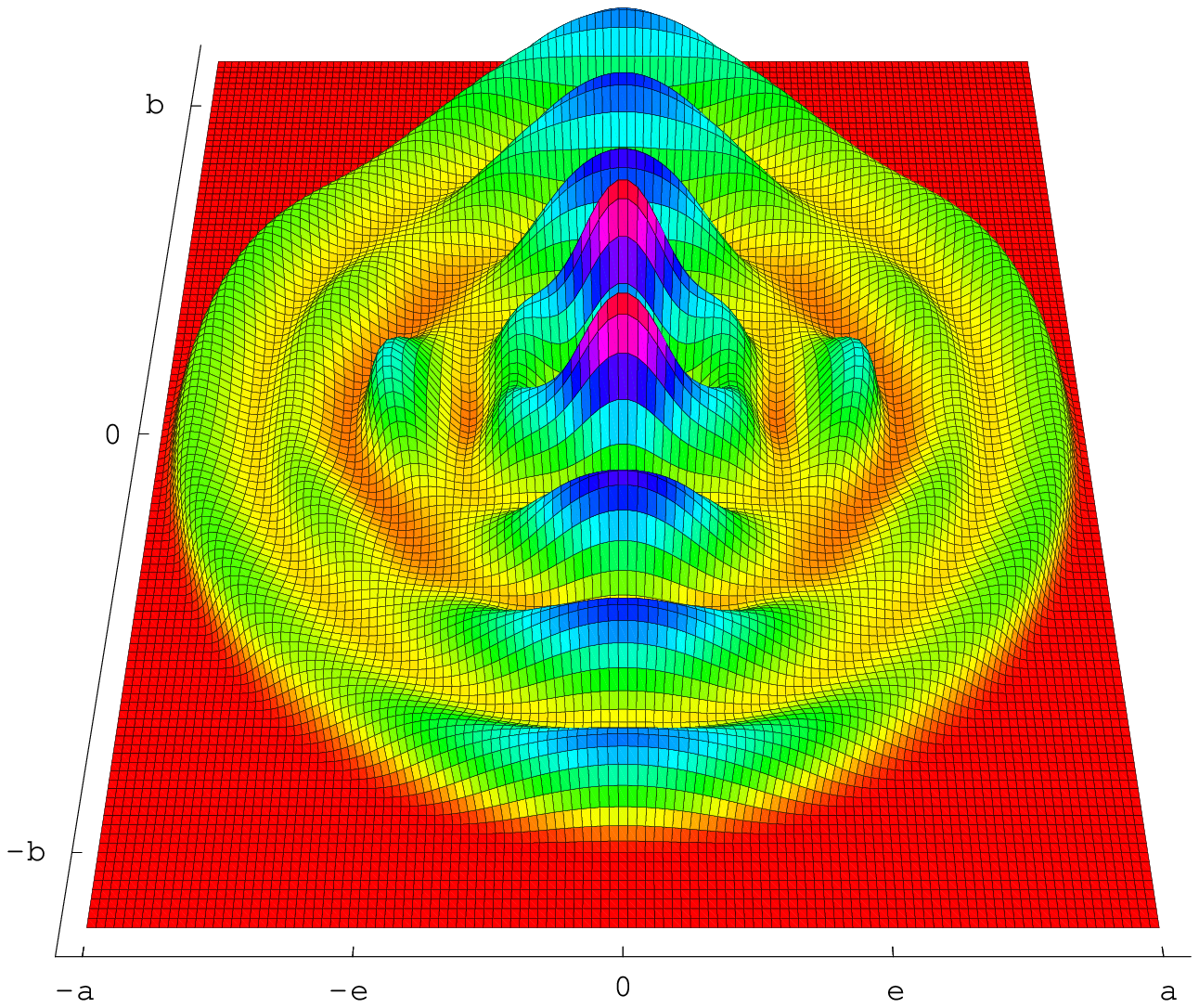}} 
\caption{LDOS $N_0({\bf r},\epsilon_{42})$, $\epsilon_{42}=\epsilon_F$, for 
non-interacting electrons inside an ellipse with semimajor axis $a$ 
(horizontal) and semiminor axis $b$ (vertical). This viewpoint is the same 
for all plots in Figs. 1-4. The scale of the LDOS can be taken from 
Figs. 5-7.}
\end{minipage}\hfill
\begin{minipage}[t]{.46\linewidth}
\epsfxsize=13cm \centerline{\epsfbox{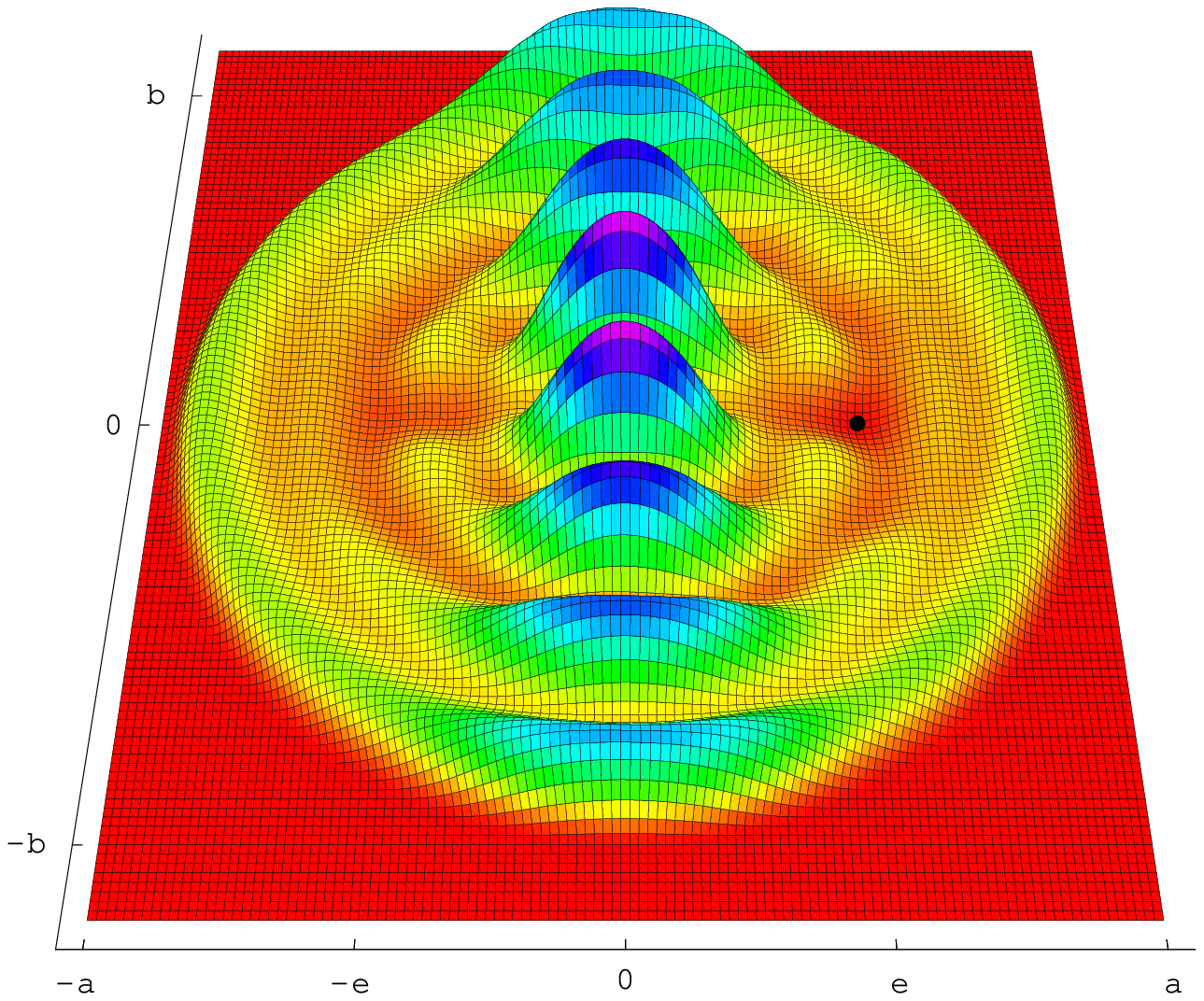}} 
\caption{LDOS $N({\bf r},\epsilon_{42})$ with an impurity ($\bullet$) at the 
right focus ${\bf r_0}=(ea,0)$; $U = 16.6 E_0A$ ($A=\pi a b$ is the area of 
the ellipse).}
\end{minipage}
\label{fig1_2}
\end{figure}
\begin{figure}[h]
\noindent
\begin{minipage}[t]{.46\linewidth}
\epsfxsize=13cm \centerline{\epsfbox{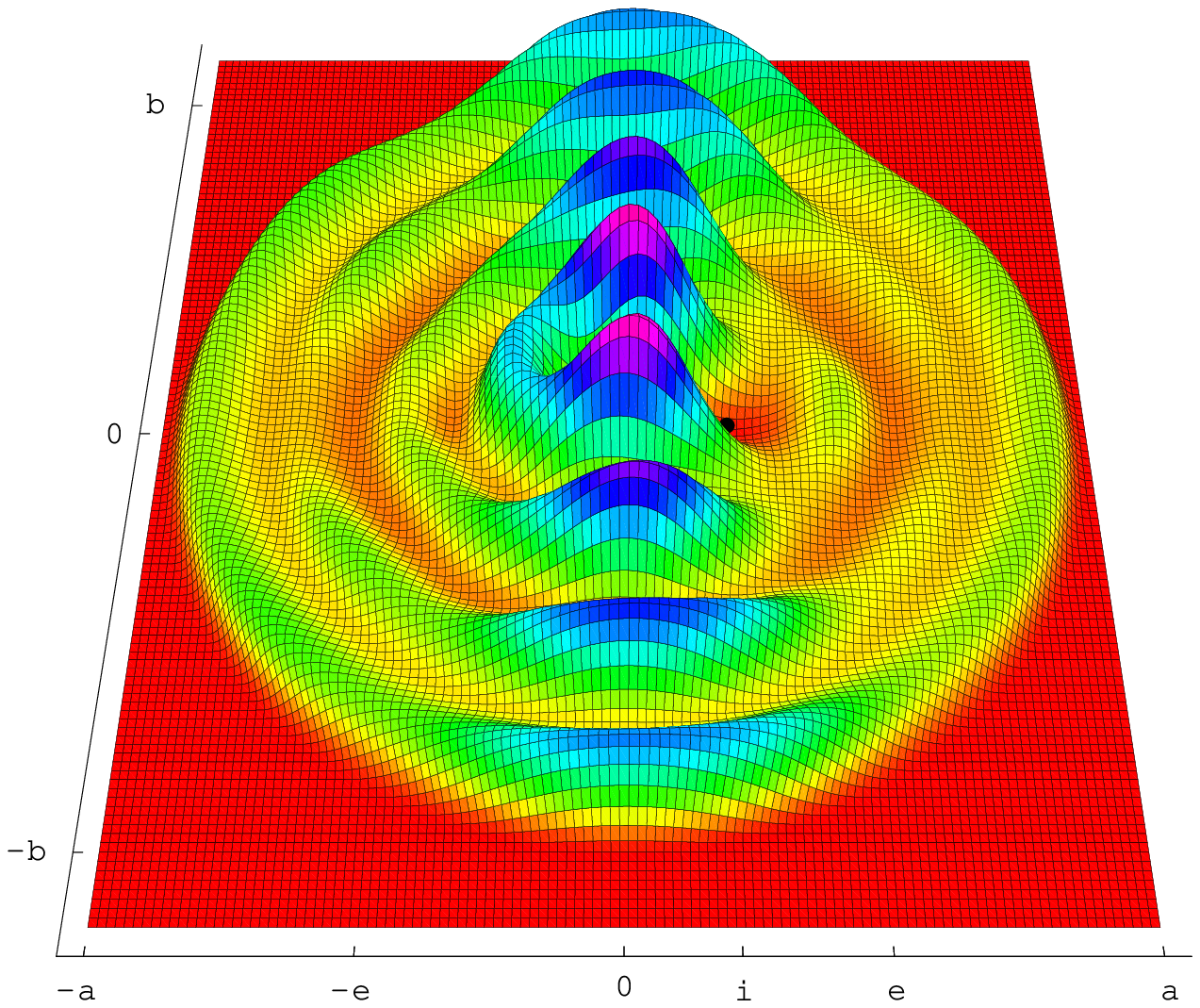}} 
\caption{LDOS $N({\bf r},\epsilon_{42})$ with an impurity ($\bullet$) at 
${\bf r_0}=(i,0)=(0.22a,0)$; $U = 16.6 E_0A$. }
\end{minipage}\hfill
\begin{minipage}[t]{.46\linewidth}
\epsfxsize=13cm \centerline{\epsfbox{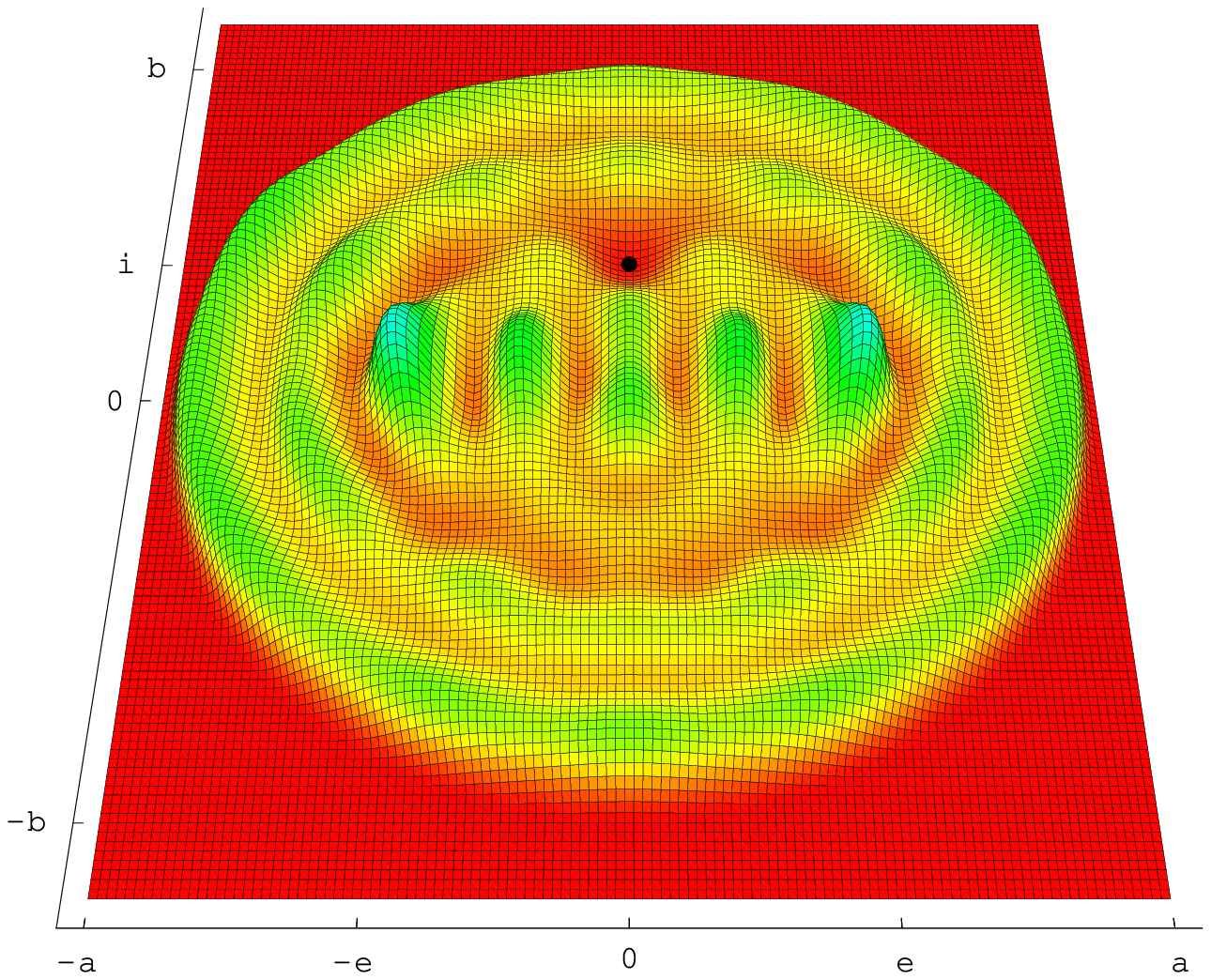}} 
\caption{LDOS $N({\bf r},\epsilon_{42})$ with an impurity ($\bullet$) at ${\bf r_0}=(i,0)=(0,0.33a)$;
$U = 16.6 E_0A$.} 
\end{minipage}
\label{fig3_4}
\end{figure}
\twocolumn

\noindent
\begin{eqnarray}
N_0({\bf r},\epsilon) = - \frac{1}{\pi} {\rm Im}\, G_0({\bf r,r},\epsilon)\, .
\end{eqnarray}
In Fig. 1 we present the LDOS at the Fermi energy $\epsilon_F=\epsilon_{42}$ 
using a finite broadening $\delta=\langle\Delta\epsilon\rangle=0.77 E_0$ 
where $\langle\Delta\epsilon\rangle$ is the average level spacing of the 
eigenenergy spectrum ($\delta=10$ meV for the parameters mentioned above for 
the ellipse with $a=71$\AA). While the eigenstates of the hard-wall ellipse 
are naturally sharp ($\delta=0$), in actual experiments the 
lifetime of the surface electrons is finite e.g. because of the imperfect 
hard-wall condition. Using a finite broadening is therefore reasonable not 
only for practical purposes. We note that 
$N_0({\bf r},\epsilon_{42})$ is dominated by the contributions of the 
42$^{nd}$ and 43$^{rd}$ state, because the energy difference between these 
states is seven times smaller than the energy difference between 41$^{st}$ 
and 42$^{nd}$ state.

Next we add a non-magnetic impurity scattering center to the ellipse at point 
${\bf r}_0$, which we model by a local delta-function potential
\begin{eqnarray}
V({\bf r})=U \delta({\bf r - r_0})\, .
\end{eqnarray}
The electronic scattering processes are thus described by the 
scattering T-matrix
\begin{eqnarray}
T({\bf r_0},\epsilon) = \frac{U}{1-U G_0^{ret}({\bf r_0,r_0},\epsilon)}\, .
\end{eqnarray}
The electronic propagator in the presence of the impurity potential follows 
then as
\begin{eqnarray}
G({\bf r, r'}, \epsilon)&=& G_0^{ret}({\bf r, r'}, \epsilon) \nonumber \\
&+& G_0^{ret}({\bf r, r_0}, 
\epsilon) T({\bf r_0}, \epsilon) G_0^{ret}({\bf r_0, r'}, \epsilon)\, ,
\end{eqnarray}
and the modified LDOS is 
\begin{eqnarray}
N({\bf r},\epsilon) = - \frac{1}{\pi} {\rm Im}\, G({\bf r,r},\epsilon)\, .
\end{eqnarray}
In our subsequent analysis we will choose selected points for the position of 
the impurity and explore the consequences for the LDOS in the entire area of
the elliptical corral.

\section{Results}
As mentioned above, Fig. 1 shows the free LDOS at the energy 
$\epsilon=\epsilon_{42}$ and for $\delta=0.77 E_0$. The contribution of the 
43$^{rd}$ state is very strong along the semiminor axis and leads to a 
sequence of local maxima and minima. On the other hand the contributions of 
the 42$^{nd}$ state are responsible for the structure of the LDOS in the 
other parts of the corral especially near the foci. 

Fig. 2 shows the LDOS when the impurity is placed at the right focus; the 
position of the impurity is indicated in the figure by a black bullet. With 
increasing 
potential strength $U$ one observes that the LDOS is reduced essentially 
everywhere except for the positions of the minima of the free LDOS and the
maxima along the semiminor axis, 
where it remains almost constant. The evolution of the LDOS with increasing 
$U$ is shown in Fig. 5 for a cut along the semimajor axis. At the 
impurity-free focus the LDOS continuously decreases with increasing $U$.
Importantly, the LDOS is almost symmetrically suppressed at the other 
impurity-free focus; this phenomenon can be ascribed to the previously studied 
so-called quantum mirage effect. Here we use this acronym to underline that 
the LDOS at both 
focal points is changing in the same manner. We emphasize again that the 
ridge structure with its high maxima along the semiminor axis is conserved.
So, although the impurity is only at the right focus, the semiminor and the 
semimajor axis still appear like symmetry axes -- at least within the 
resolution of the chosen color grid. This is the quantum mirage.
A similar mirage effect persists also, when the impurity is moved slightly 
away from the focus \cite{Lobos}.

\begin{figure}[h]
\centering\epsfig{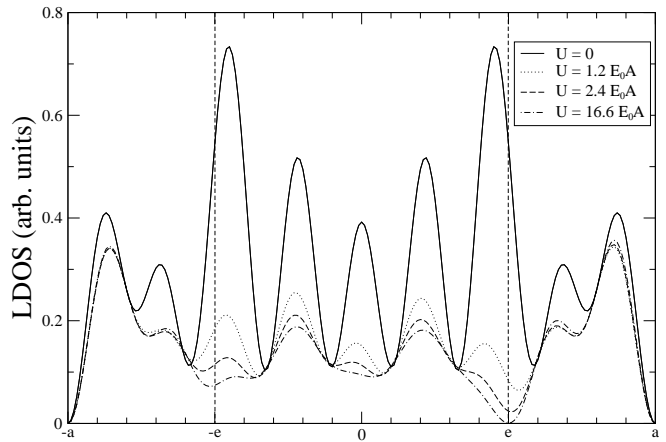}
\vskip0.5cm
\caption{LDOS $N({\bf r},\epsilon)$ for a cut along the semimajor axis; the 
impurity is at the right focus. The vertical dashed lines serve as a guide 
to the eye to compare the LDOS at the impurity with the LDOS at the mirror 
position.}
\label{fig5}
\end{figure}

A very different behavior of the LDOS is observed, if the impurity is placed 
at the position of the first maximum of the free LDOS on the semimajor axis 
near the center of the ellipse. This is shown in Fig. 3, and the corresponding 
cut along the semimajor axis is shown in Fig. 6. Again, as expected, the 
LDOS in the presence of the impurity is reduced with increasing $U$ almost 
everywhere inside the corral. But, surprisingly, the opposite effect is 
observed at the mirror image point of the impurity position with respect to a 
reflection at the origin. There the impurity causes in fact an enhancement of 
the LDOS. Thus, in this case the interference pattern for the perturbed 
electronic wavefunctions leads just to the opposite effect as in the above 
discussed quantum mirage in Fig. 2; we therefore call this observation an 
anti-mirage effect. Note that in the rest of the corral the shape of the 
impurity LDOS remains very symmetric, but with a slight overall enhancement 
of the LDOS in the left part of the ellipse. Similar as in Fig. 2 the maxima 
of the ridge structure along the semiminor axis persist almost unchanged.
One can observe the anti-mirage effect also when the impurity is put 
somewhere between ${\bf r_0}=(0.16a,0)$ and the second minimum at
 ${\bf r_0}=(0.33a, 0)$.
\begin{figure}[h]
\centering\epsfig{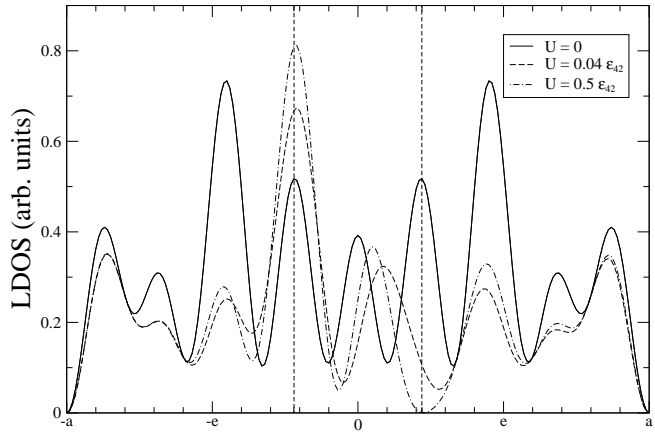}
\vskip0.5cm
\caption{LDOS $N({\bf r},\epsilon_{42})$ for a cut along the semimajor axis 
with the impurity at the position ${\bf r_0}=(0.22a,0)$.}
\label{fig6}
\end{figure}

In a third example we place the impurity at the second maximum along the 
semiminor axis away from the center (see Fig. 4 and Fig. 7). This particular
 impurity 
position has the remarkable consequence that the ridge structure along 
the semiminor axis is entirely wiped out, and what is left is very similar to 
the probability density of the 42$^{nd}$ state. This phenomenon occurs for 
every impurity position at any of the local maxima along the semiminor axis. 
For example, if the impurity is placed at 
the points with the highest LDOSs of the elliptic corral, the LDOS reacts most 
sensitively. This in itself may not appear as a surprise, because a local 
perturbation at the space point with the highest probability density should 
indeed lead to severe changes in the local electronic structure. Yet, the 
total wipeout of the ridges comes as a spectacular surprise. If the impurity 
is placed at a minimum along the semiminor axis, the resulting LDOS appears 
essentially unaffected.

We note that the LDOS strongly depends on the electronic density in the corral,
i.e. the spatial structure of the eigenstates at or close to the Fermi energy.
Different situations can be realized by varying the length of the semimajor 
axis at fixed electronic density. They can be physically achieved by studying
different corral sizes on identical substrates. For example, the energies of 
the $42^{nd}$ and the $43^{rd}$ state are very close to each other so that 
the LDOSs with the same broadening $\delta$ at the eigenenergies 
$\epsilon_{42}$ and $\epsilon_{43}$ are nearly the same. On the other hand, 
the LDOSs at energies $\epsilon_{41}$ and $\epsilon_{44}$ have a 
significantly different structure. In particular, the LDOSs 
$N({\bf r},\epsilon_{41})$ and $N({\bf r},\epsilon_{44})$ nearly vanish at
the foci. The quantum mirage effects as arising from placing impurities near 
the foci, are in this case essentially unobservable. In this sense, the size 
of the corrals of Co atoms as realized in the experiments was a lucky choice 
for detecting the mirage phenomena.  

\begin{figure}[h]
\centering\epsfig{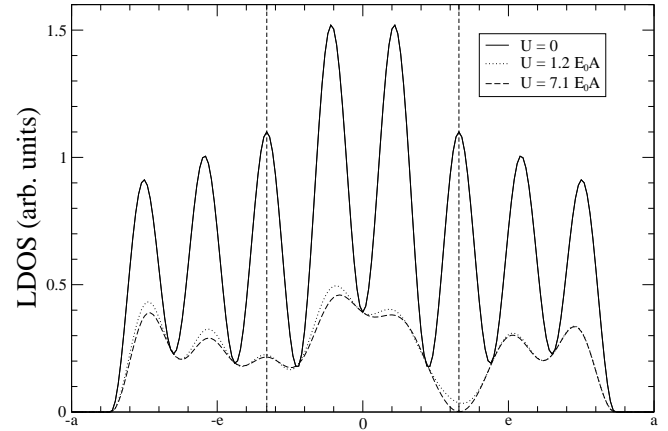}
\vskip0.5cm
\caption{LDOS $N({\bf r},\epsilon_{42})$ for a cut along the semiminor axis 
with the impurity at ${\bf r_0}=(0,0.33a)$.}
\label{fig7}
\end{figure}

\section{SUMMARY}
We have identified three distinctly different interference phenomena for 
non-magnetic impurity induced changes in the LDOS of non-interacting 
electrons in elliptic hard-wall quantum corrals. Mirage or anti-mirage effects 
occur depending on the impurity position inside the corral. Even an almost 
complete suppression of pronounced rich structures in the LDOS can be achieved 
for special choices of the impurity location. These surprising phenomena
are a manifestation of quantum mechanical interference effects which may be
tested experimentally with the already existing elliptic quantum corrals
on metallic surfaces.

\section*{ACKNOWLEDGMENTS}
We thank M. Sekania, R. Bulla, N. Tong, and X. Ren for assistance
and helpful discussions. This work was supported by the Deutsche 
Forschungsgemeinschaft through SFB 484.


\begin{references}
\bibitem{Eigler} D.~M. Eigler and E.~K. Schweitzer, Nature {\bfseries344}, 524 
(1990).
\bibitem{Fiete} For a recent review see G.~A. Fiete and E.~J. Heller, Rev. 
Mod. Phys. {\bf 75}, 933 (2003).
\bibitem{Manoharan} H.~C. Manoharan, C.~P. Lutz, and D.~M. Eigler, Nature 
{\bfseries403}, 512 (2000).
\bibitem{Schiller} O. Agam and A. Schiller, Phys. Rev. Lett. {\bfseries86}, 484
(2001).
\bibitem{Lobos} A. Lobos and A.~A. Aligia, Phys. Rev. B {\bf 68}, 035411 
(2003).  
\bibitem{Aligia} A.~A. Aligia, Phys. Rev. B {\bfseries64}, 121102 (2001).
\bibitem{Chiappe} G. Chiappe and A.~A. Aligia, Phys. Rev. B {\bfseries66},
 075421 (2002).
\bibitem{Porras} D. Porras, J. Fern\'{a}ndez-Rossier, and C. Tejedor, 
Phys. Rev. B {\bfseries63}, 155406 (2001).
\bibitem{McLachlan} N.~W. McLachlan, Theory and Applications of Mathieu
Functions (Dover Publications, New York, 1964).
\bibitem{Stegun} M. Abramowitz and I. A. Stegun, Handbook of 
Mathematical Functions (Dover Publications, New York, 1965). 

\end{references}
\end{document}